\documentclass[aps,pre,reprint,groupedaddress]{revtex4-1}

\usepackage{graphicx}
\usepackage{dcolumn}
\usepackage{bm}
\usepackage{float}
\usepackage{amsmath}
\usepackage{verbatim}
\usepackage[utf8]{inputenc}

\DeclareMathOperator\arctanh{arctanh}

\begin{document}

\title{Random Sequential Adsorption on mobile patches }

\author{Diogo E. P. Pinto and Nuno A. M. Ara\'{u}jo}
\affiliation{Departamento de F\'{i}sica, Faculdade de Ci\^{e}ncias, and Centro de F\'{i}sica Te\'{o}rica e Computacional, Universidade de Lisboa, Campo Grande, P-1749-016 Lisboa, Portugal}

\date{\today}

\begin{abstract}

An extension of the Random Sequential Adsorption (RSA) model has been proposed recently, motivated by the coverage of oil droplets by DNA-functionalized colloidal particles. Particles arrive to a flat substrate with a uniform flux $F$ but they can only adsorb on patches. Patches diffuse on the substrate with a diffusion coefficient $D$ if they are free and they remain immobile when attached to an adsorbed particle. The adsorption is considered irreversible and particles cannot adsorb on top of each other. Thus, the system reaches a jammed state, consisting of a monolayer where no more particles can adsorb. We performed Monte Carlo simulations to study the adsorption kinetics and jammed-state morphology on a one-dimensional lattice. We show that, while the time-dependence of the coverage depends on $F$ and $D$, the jammed-state coverage depends solely on the ratio $F/D$. This result is grasped by a simple mean-field calculation. We also report two different regimes for the functional dependence of the jammed-state coverage on the size of the particles, for low and high density of patches.
\end{abstract}


\maketitle

\section{Introduction}

Surface adsorption has been a broadly researched topic over the last decades \cite{Privman1994, Kumacheva2002, Fustin2003, Biancaniello2005, Cadilhe2007, Wang2012, Oliveira2014, Araujo2016}. For practitioners, particle adsorption on substrates enables a wide range of applications ranging from photonic crystals, to quantum dots, sensors, and encapsulation \cite{Boerasu2002, Burda2005, Joshi2016}. Theoretically, understanding the kinetics of adsorption poses fundamental challenges to non-equilibrium statistical physics \cite{Evans1993, Bartelt1991b, Talbot2000}. In the limit of irreversible adsorption, the prototypical model is the Random Sequential Adsorption (RSA) model, where adsorption is considered irreversible and particle-particle interactions are excluded volume. Despite simple, RSA provides valuable information about the adsorption kinetics and morphology of the final structure \cite{Evans1993, Cadilhe2007, Privman2000a}.

Different extensions of RSA were proposed to study the role of the particle size \cite{Bartelt91c, Bonnier2001a}, particle shape \cite{Pomeau1980, Swendsen1981, Dias2012}, and particle-particle correlations \cite{Evans1993}. With experimental techniques reaching smaller and smaller length scales came the possibility of engineering substrates featuring patterns in the length scale of the particle size. This paved the way to new experiments and theoretical studies on how to control the morphology of the final structure, using patterns that interact selectively with the particles \cite{Chen2002, Kumacheva2002, Dziomkina2005, Joo2006, Cadilhe2007, Araujo2008, Kelzenberg2010, Marques2012, Katsman2013a, Katsman2013b, Privman2016, Ciesla2017}. So far, for simplicity, these patterns were considered static.

\begin{figure*}[t!]
	\includegraphics[scale=0.6]{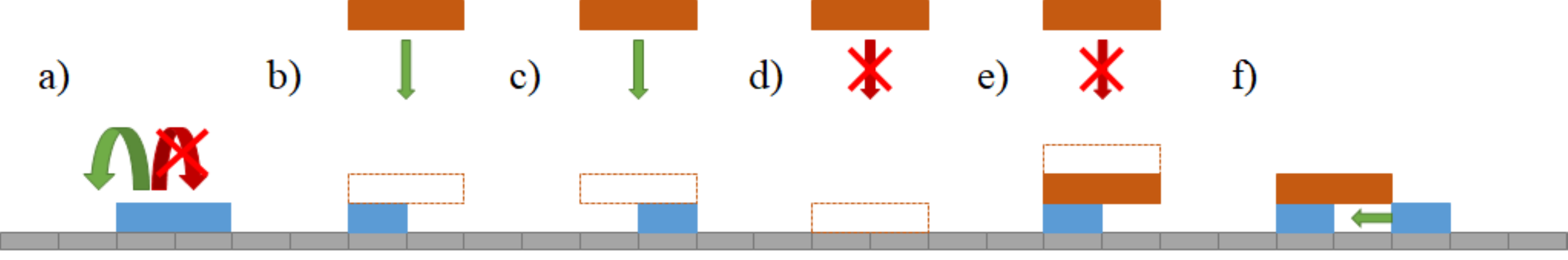}
	\caption{\label{Figura1}Schematic representation of the model. (a) Patches (in blue) occupy one site and diffuse on the lattice when they are free. The patch-patch interaction is excluded volume. Particles (in orange) can only adsorb on free patches (b, c, and d) and cannot overlap any previously adsorbed particle (e). A free patch can diffuse underneath a previously adsorbed particle, binding to it (f).}
\end{figure*}

Recently, an experimental protocol was proposed where colloidal particles adsorb irreversibly on the surface of an oil droplet, using DNA-functionalization. The surface of the oil droplet is covered with patches that diffuse on the surface and act as landing sites for the DNA-functionalized colloidal particles. By contrast to particles at interfaces, when the coverage of the surface is mediated by the patches, the strong capillary forces are suppressed. An extension of RSA was proposed to help explaining the experimental results \cite{Joshi2016}. In the model, particles of a certain size attempt adsorption on a substrate with mobile patches. The particles only adsorb on free patches and adsorption is irreversible. A free patch can find a particle adsorbed previously and bind to it. Here, we discuss the kinetics of adsorption and morphology of the final structure for this model. 

The paper is organized as follows. The model is described in Section II. In Section III, we study how the competition between timescales affects the kinetics of adsorption and we develop a mean-field calculation that sheds light on the numerical results. We discuss also the morphology of the jammed state. Some conclusions are drawn in Section IV.

\section{Model}

Figure \ref{Figura1} shows a schematic representation of the model, first proposed in Ref. \cite{Joshi2016}. The substrate is described as a one-dimensional lattice with $L$ sites. A fraction $n_{0}$ of the sites is occupied by patches of size one (blue squares). Free patches diffuse with a diffusion coefficient $D$, defined as the rate at which each patch hops to one of its two first neighbors. We consider a patch-patch excluded volume interaction, i.e., a patch cannot move into an occupied site (Fig. \ref{Figura1}a).

Particles are discrete segments of size $ k $, in units of lattice sites (orange rectangles in Fig. \ref{Figura1}). They arrive sequentially to the substrate at random positions with a flux $ F $, defined as the rate of adsorption attempts per unit time and length. If a particle attempts adsorption on a free patch and does not overlap a previously adsorbed particle, the adsorption is successful, forming a particle-patch complex (Figs. \ref{Figura1}b and c) and adsorption fails otherwise (Figs. \ref{Figura1}d and e). The particle-patch complexes are immobile and the adsorption is irreversible, i.e., an adsorbed particle cannot detach from the patch.

For $ k=1 $, the number of adsorbed particles per lattice site saturates asymptotically at $ n_{0} $. Hereafter, we consider the non-trivial case of $ k>1 $. For $ k>1 $, a free patch can diffuse and go underneath a previously adsorbed particle, provided that it does not overlap any other patch (Fig. \ref{Figura1}f). In this case, the patch binds irreversibly to the adsorbed particle. The standard RSA model is recovered in the limit where all sites are occupied by patches ($n_0=1$).

We performed kinetic Monte Carlo simulations considering two processes: adsorption attempts at a rate $ F $ and free-patch diffusion at a rate $ D $. At each iteration, time is incremented by:

\begin{equation}
\Delta t=-(n_{F}F+n_{D}D)^{-1}\log(r),
\end{equation}   

\noindent where r is a random number distributed uniformly in the range $ ]0,1] $, and $ n_{F} $ and $n_D$ are the number of possible adsorption and diffusion events, respectively. We performed simulations on a lattice of length $ L=10^{6} $ and results are averaged over $ 10^{4} $ samples. We considered also other system sizes and concluded that, for the considered value of $L$, finite-size effects are negligible.

\section{Results}

We first study the case of dimers ($ k=2 $) and then proceed analyzing the effect of the particle size, considering $ k $-mers ($ k>2 $). We also discuss the dependence on the initial density of patches.

\subsection{Adsorption of dimers}

\begin{figure}[b!]
	\includegraphics[scale=0.5]{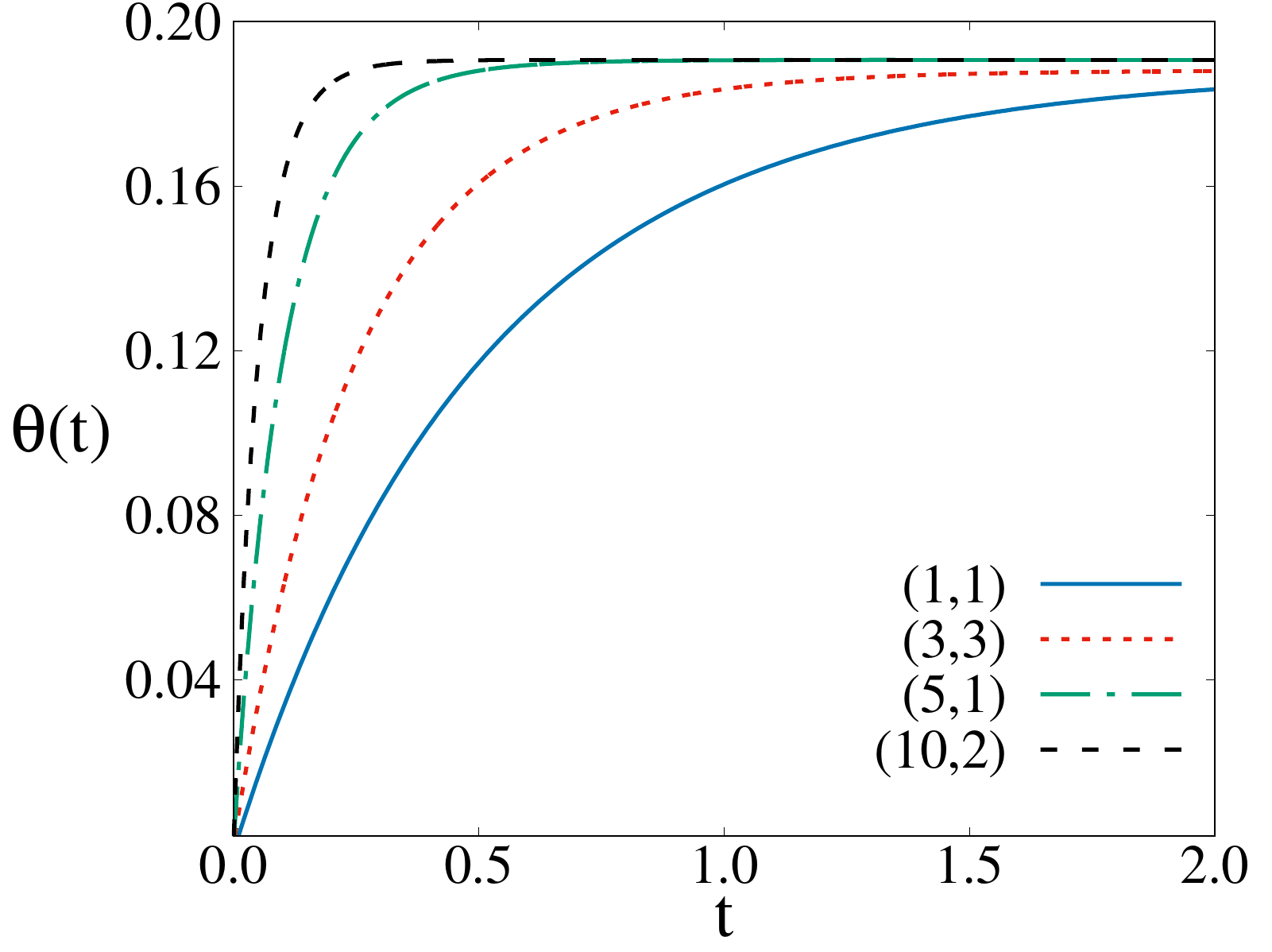}
	\caption{\label{Figura3}Time-dependence of the coverage for $ (F,D)=\{(1,1),(3,3),(10,2),(5,1)\} $ and $ n_{0}=0.1 $. Results are averages over $10^{4}$ samples of a one-dimensional lattice with $L=10^{6}$ sites.}
\end{figure}

\begin{figure}[t!]
	\includegraphics[scale=0.5]{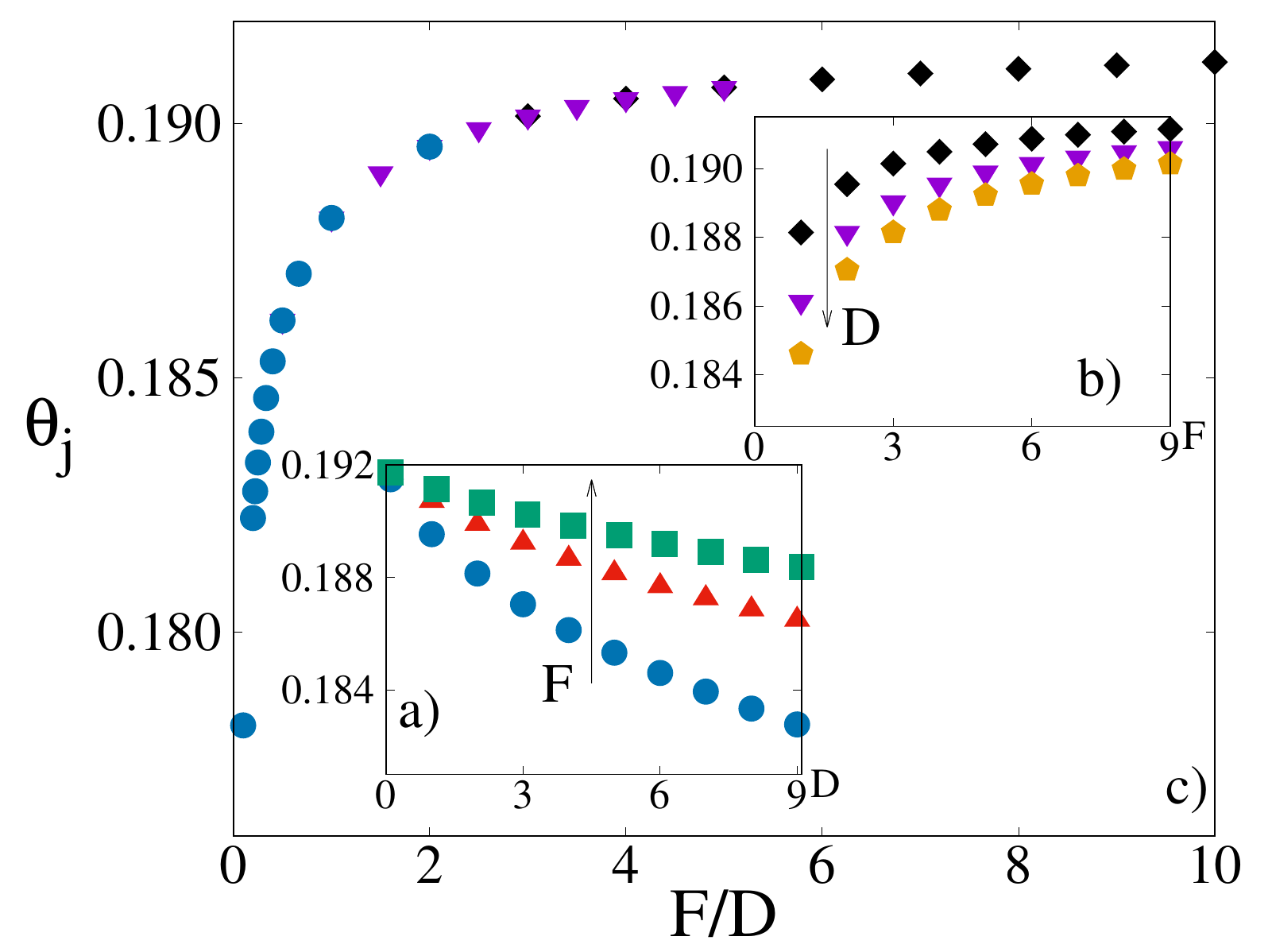}
	\caption{\label{Figura4} Jammed-state coverage as a function of a) $ D $ for $ F=\{2,5,10\} $ and $ n_{0}=0.1 $, represented as circles, triangles, and squares, respectively, b) $ F $ for $ D=\{1,2,3\} $ and $ n_{0}=0.1 $, represented as pentagons, rhombus, and inverted triangles, respectively, and c) $ F/D $ for three values of $ F $ and $ D $, considered in a) and b). Results are averages over $10^{4}$ samples of a one-dimensional lattice with $L=10^{6}$ sites.}
\end{figure}

\begin{figure}[b!]
	\includegraphics[scale=0.5]{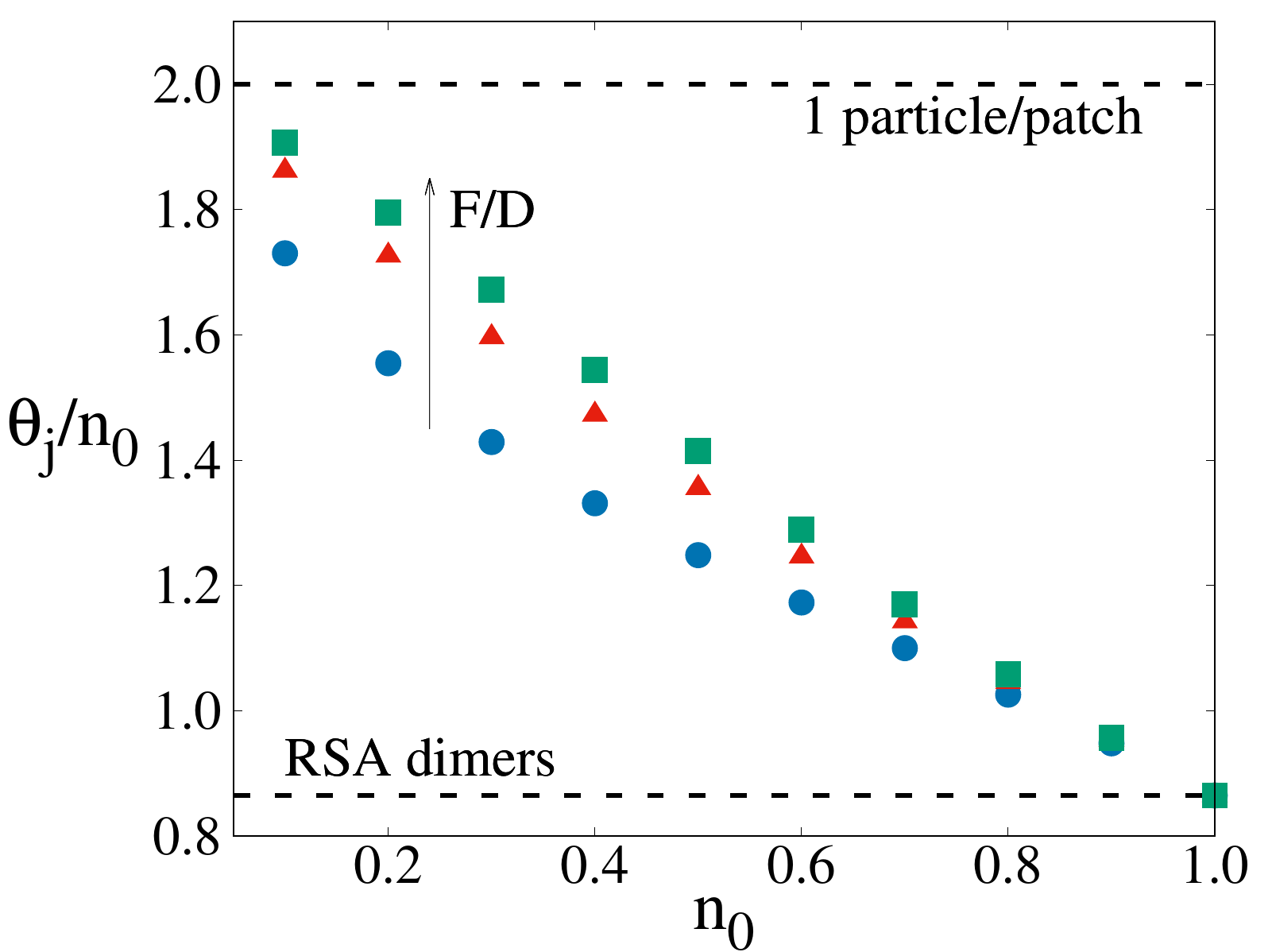}
	\caption{\label{Figura9} Jammed-state coverage per patch as a function of $ n_{0} $ for different $ F/D $. As $ n_{0} $ increases, $ \theta_{j}/n_{0} $ converges to the RSA of dimers, $ \theta_{j}(n_{0}=1)=1-e^{-2} $ \cite{Flory1939}. Results are averages over $10^{4}$ samples of a one-dimensional lattice with $L=10^{6}$ sites.}
\end{figure}

The adsorption is irreversible and particles can only adsorb on patches. We define coverage as $ \theta=kN/L $, where $ N  $ is the number of adsorbed particles. The coverage is expected to monotonically increase until it saturates for the jammed state, where no more particles can adsorb. In the jammed state there might be some free patches, but they are trapped inside gaps that are smaller than the particle size and, therefore, no particle can adsorb on them. We define $\theta_j$ as the jammed-state coverage.

For very low density of patches ($ n_{0} $), the average distance between patches is such that the adsorption on different patches can be decoupled. In such limit, there will be only one particle per patch in the jammed state and, thus, $\theta_{j}=kn_{0}$. This value sets also an upper bound for the coverage, as it is not possible to have more particles adsorbed than patches. 

Figure \ref{Figura3} shows the time evolution for the coverage, for different values of the flux $F$ and diffusion coefficient $D$. The coverage initially increases and saturates asymptotically at $\theta=\theta_{j}$. The kinetics evolves as a competition between two mechanisms: adsorption on and diffusion of free patches. The former occurs with an inter-arrival time $\tau_{F}\propto 1/F$ and the latter occurs in a timescale $\tau_{D}\propto 1/D$. If $\tau_{D}\ll\tau_{F}$, diffusion is much faster than adsorption and thus, in between adsorption events, patches typically have enough time to diffuse and go underneath particles that adsorbed previously (Fig. \ref{Figura1}f). When $\tau_D\gg\tau_F$, diffusion can be neglected and the coverage is maximized. Thus, for constant $F$, the jammed-state coverage $\theta_{j}$ decays with $D$, as shown in Fig. \ref{Figura4}a. In the same way, $\theta_{j}$ increases with $F$ for constant $D$ (see Fig. \ref{Figura4}b). Numerically, a data collapse is obtained when the jammed-state coverage is plotted as a function of $F/D$, as shown in Fig. \ref{Figura4}c. Note that, while the jammed-state coverage solely depends on $F/D$, the kinetics towards the jammed state depends on $F$ and $D$ independently (see Fig. \ref{Figura3}).

The kinetics also depends on the concentration of patches $n_{0}$. To go underneath a previously adsorbed particle (Fig. \ref{Figura1}f), a free patch needs to diffuse over a distance that corresponds to the average separation between particles. The lower the value of $n_{0}$ the larger is that distance. Thus, the number of particles per patch in the jammed state, given by $\theta_{j}/n_{0}$, decays monotonically with $n_{0}$, as shown in Fig. \ref{Figura9} for different ratios of $F/D$. For $n_{0}=1$, every site is occupied by a patch and the results for standard RSA are recovered, independently of $F/D$.

\begin{figure}[t!]
	\includegraphics[scale=0.5]{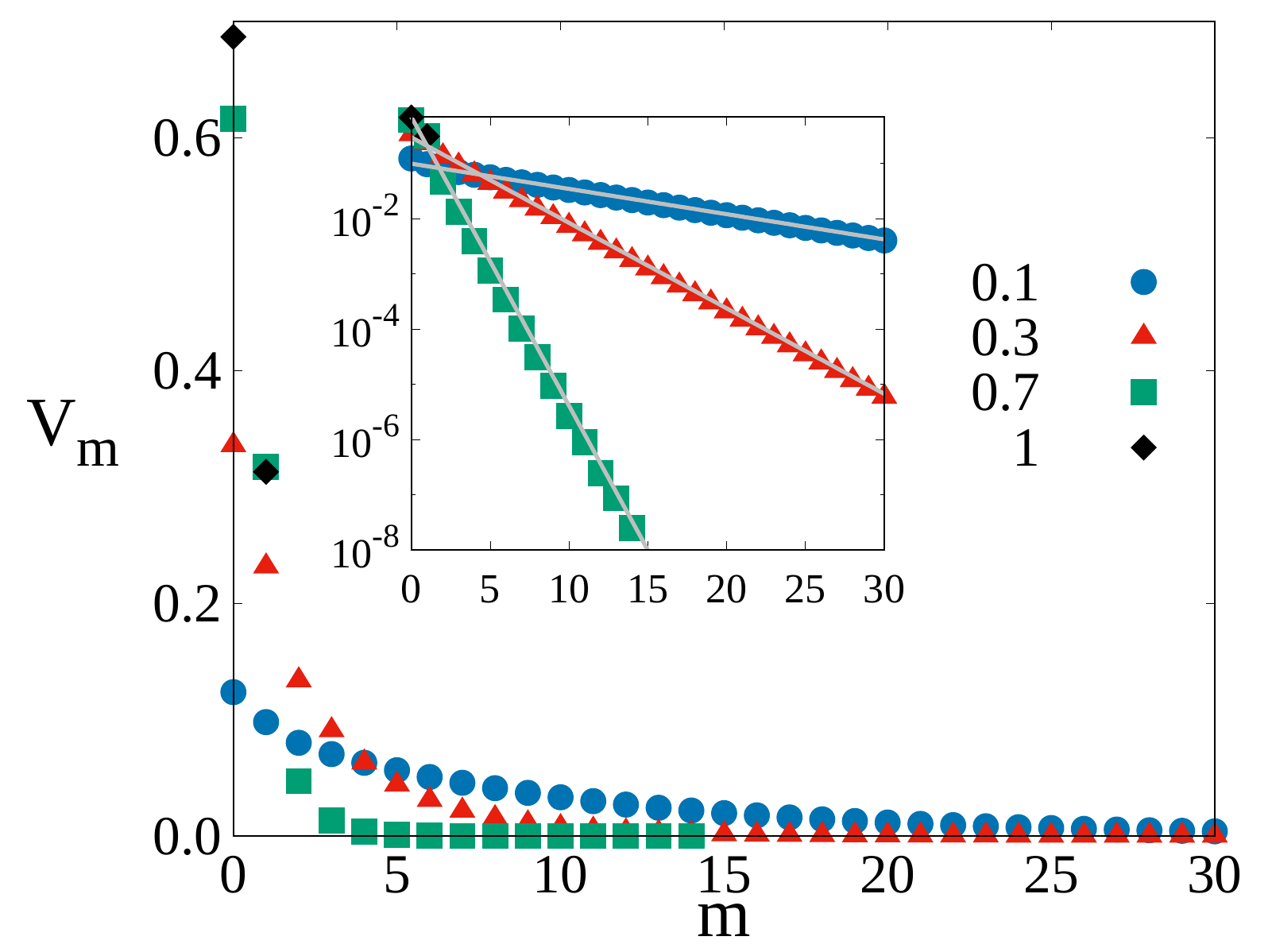}
	\caption{\label{Figura10} Gap-size distribution function at the jammed state for $ n_{0}=\{0.1, 0.5, 0.7, 1.0\} $ (semi-log plot in the inset). In the inset, the three curves are given by $ V_{m}=n_{0}(1-n_{0})^{m} $, with $ n_{0}=\{0.1, 0.5, 0.7\} $. Results are averages over $10^{4}$ samples of a one-dimensional lattice with $L=10^{6}$ sites.}
\end{figure}

To characterize the morphology of the jammed state, we measured the gap-size distribution function ($ V_{m} $), defined as the probability of finding a sequence of $m$ sites not occupied by particles (gap) in between two particles. Figure \ref{Figura10} shows $ V_{m} $ for different $ n_{0} $. For RSA ($ n_{0}=1 $), all gaps are smaller than the particle size. However, this is not the case for $ n_{0}<1 $. In this case, gaps of size $m>k$ are possible, provided that there is no free patch inside the gap. 

The probability of finding a gap $m>k$ decays exponentially with $m$ (Fig. \ref{Figura10}). This can be explained as follows. For simplicity, let us consider the adsorption of monomers. There will be one adsorption per patch and no particle-particle correlations. Thus, the jammed state consists of a lattice with a fraction $n_{0}$ of sites occupied at random. The gap-size distribution function is then $ V_m=n_{0}^{2}(1-n_{0})^{m} $.
Since $ (1-n_{0})<1 $, $ V_{m} $ decays exponentially with $m$. By fitting the data in the inset of Fig. \ref{Figura10}, for $m>k$, with a function $f(m)=a.b^m$, we obtain $b\approx1-n_0$, in line with the predicted $V_m$. 

\subsection{Mean-field approach}

Let us now consider a mean-field approach, based on rate equations. We consider a uniform distribution of patches and neglect particle-particle correlations. We define $\rho_{r}$ and $\rho_{p}$ as the density of free patches and adsorbed particles, respectively. Initially, $\rho_{r}(0)=n_{0}$ and $\rho_{p}(0)=0$. The coverage at every time is given by $\theta(t)=k\rho_{p}(t)$, where $k$ is the particle size.

The kinetics can be described by the following rate equations,

\begin{equation}
\label{initial}
\begin{cases}
\dot{\rho_{r}}(t)=-\overline{F}\rho_{r}-\overline{D}\rho_{p}\rho_{r} \  \\
\hspace{5cm} , \\ 
\dot{\rho_{p}}(t)=\overline{F}\rho_{r} \ 
\end{cases}
\end{equation}

\noindent where $ \overline{F} $ and $ \overline{D} $ are monotonic increasing functions of the flux $ F $ and the diffusion coefficient $ D $, respectively. In the first equation, the first term on the right-hand side corresponds to the adsorption of particles on free patches (Figs. \ref{Figura1}b and c). The second term is related to free patches going underneath previously adsorbed particles (Fig. \ref{Figura1}f).  In the second equation, the density of particles is only affected by adsorption events (gain term). 

This system of equations can be solved exactly for the considered boundary conditions,

\begin{widetext}
\begin{equation}
\label{finalrohr}
\rho_{r}(t)=\frac{\overline{F}+2\overline{D}n_{0}}{\overline{D}+\overline{D}\cosh\bigg[\sqrt{\overline{F}(\overline{F}+2\overline{D}n_{0})} t+2\arctanh\bigg(\sqrt{\frac{\overline{F}}{\overline{F}+2\overline{D}n_{0}}}\bigg)\bigg]} \ ,
\end{equation}

\begin{equation}
\label{finalrohp}
\rho_{p}(t)=\frac{\sqrt{2\overline{F}\overline{D}n_{0}+\overline{F}^{2}}\tanh\bigg\{\frac{1}{2}\bigg[2\arctanh\bigg(\frac{\sqrt{\overline{F}}}{\sqrt{\overline{F}+2\overline{D}n_{0}}}\bigg)+\sqrt{\overline{F}}\sqrt{2\overline{D}n_{0}+\overline{F}}t\bigg]\bigg\}-\overline{F}}{\overline{D}} \ .
\end{equation}
\end{widetext}

The jammed state can be obtained taking the asymptotic limit of Eqs. (\ref{finalrohr}) and (\ref{finalrohp}),

\begin{equation}
\label{assrohr}
\rho_{r}(\infty)=0 \ ,
\end{equation}

\noindent and,

\begin{equation}
\label{assrohp}
\rho_{p}(\infty)=\sqrt{\overline{F}/\overline{D}}\sqrt{2n_{0}+\overline{F}/\overline{D}}-\overline{F}/\overline{D} \ .
\end{equation}

\noindent Equation (\ref{assrohp}) predicts that the jammed-state coverage depends only on $ \overline{F}/\overline{D} $, although for the time evolution, given by Eq. (\ref{finalrohp}), this rescaling is not possible. This is consistent with what was observed numerically (see Figs. \ref{Figura3} and \ref{Figura4}). 

\subsection{Adsorption of $ k $-mers}

\begin{figure}[b!]
	\includegraphics[scale=0.5]{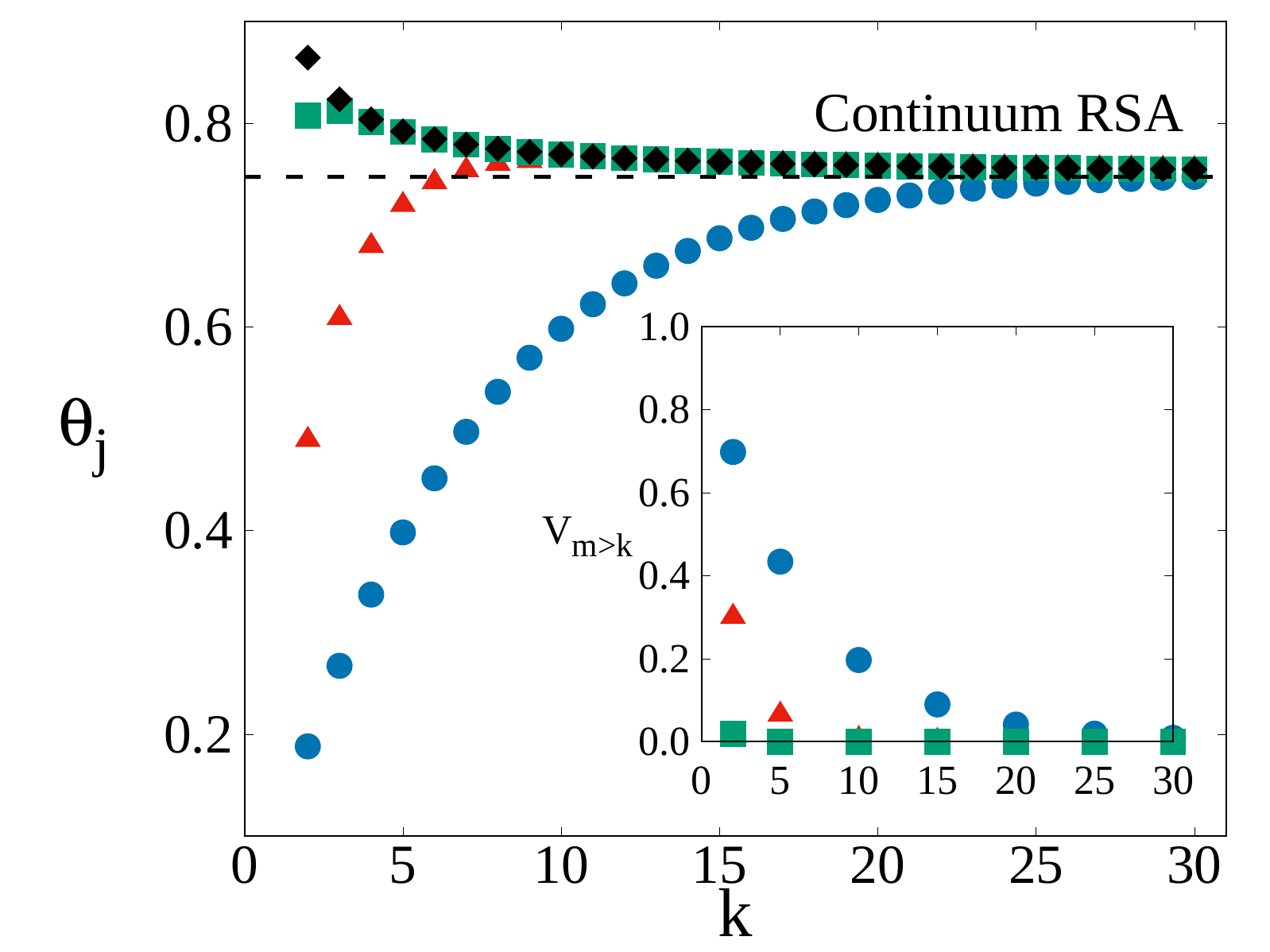}
	\caption{\label{Figura12} Jammed-state coverage as a function of the particle size for different initial density of patches, $n_{0}=\{0.1,0.3,0.7,1.0\}$ (symbols as in Fig. \ref{Figura10}) and $ F=D=1 $. The inset shows the probability of finding gaps larger than the particle size, $ V_{m>k} $, for different $ k $. For high values of $ k $, the system converges to the continuum limit of RSA (dashed line) \cite{Renyi1958}. Results are averages over $10^{4}$ samples of a one-dimensional lattice with $L=10^{6}$ sites.}
\end{figure}

We now consider the effect of the particle size. Figure \ref{Figura12} shows the jammed-state coverage as a function of the particle size ($ k $) for different $ n_{0} $. For RSA, the coverage monotonically decreases with the particle size. For larger particles, the gaps where no more particles can adsorb are also large. Consequently, the average gap size will always increase with the particle size. By contrast, for the model considered here, we find a range of parameters where the coverage increases with $ k $. 

\begin{figure}[t!]
	\includegraphics[scale=0.5]{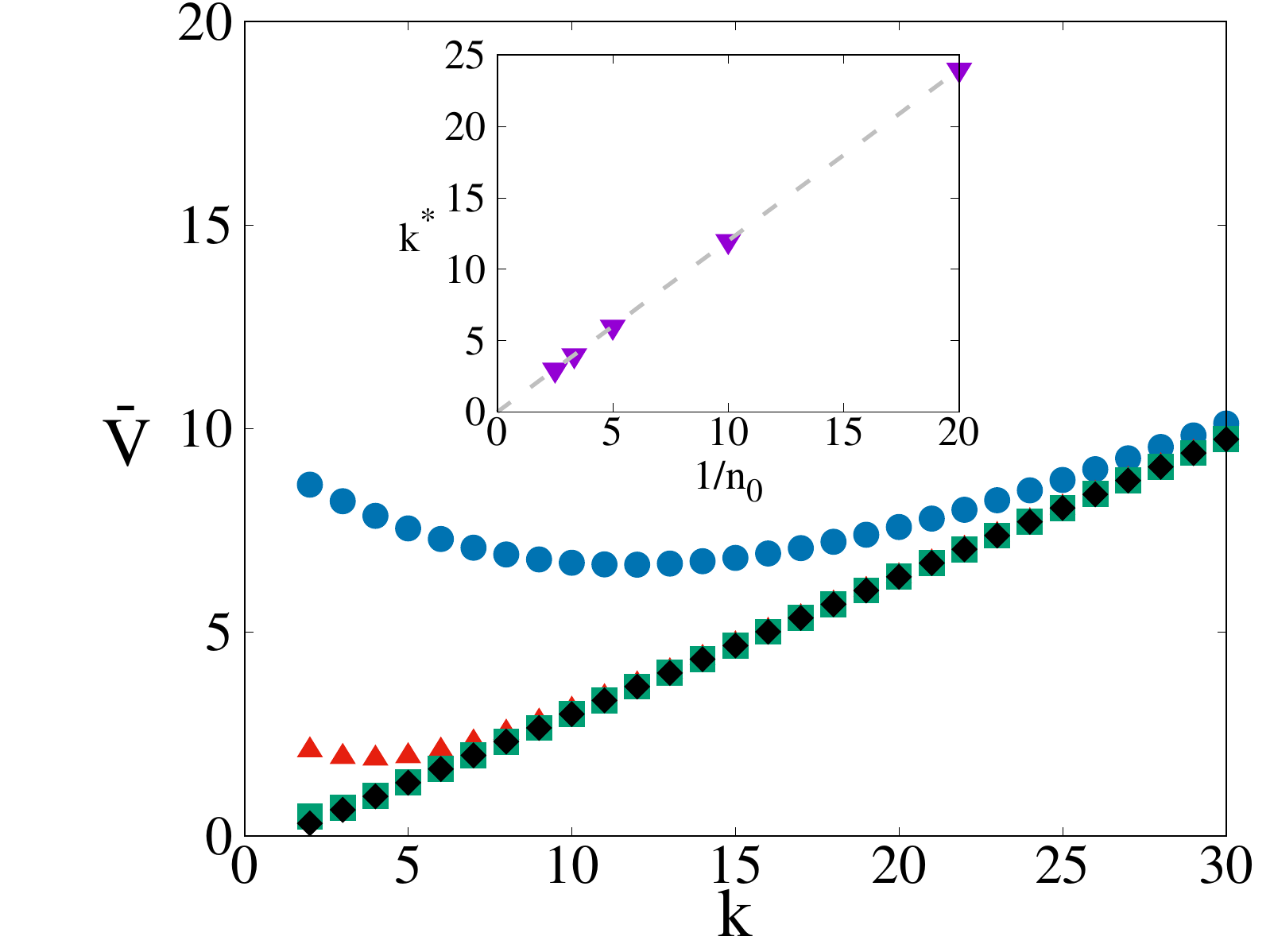}
	\caption{\label{Figura14} Average gap size as a function of the particle size. Symbols correspond to $n_{0}=\{0.1,0.3,0.7,1.0\}$ (symbols as in Fig. \ref{Figura10}). In the inset is the value of $k$ corresponding to the minimum as a function of $ 1/n_{0} $. Results are averages over $10^{4}$ samples of a one-dimensional lattice with $L=10^{6}$ sites.}
\end{figure}

To shed light on the increase of the coverage with $k$, observed for low values of $n_{0}$, we consider the dependence on $k$ of the average gap size, defined as $ \overline{V}=\sum_{m=0}^{\infty}mV_{m} $ (Fig. \ref{Figura14}). For large values of $k$ all curves converge to the one expected for RSA ($n_{0}=1$). While, for large values of $n_{0}$ the average gap size increases with $k$, for  $n_{0}= 0.1$ and $0.3$, we find an optimal value of $k=k^{*}$ at which the average gap size is minimized. For $ k\ll l $, where $l=1/n_{0}$ is the average distance between free patches, adsorption on different patches can be decoupled and the average number of adsorbed particles converges to the number of patches. In this regime, the distance between adsorbed particles is expected to decrease with $k$. Thus, $k^{*}$ sets the size above which the particle-particle correlations developed during the adsorption process become relevant. Accordingly, $k^{*}$ scales linearly with $1/n_{0}$, as shown in the inset of Fig. \ref{Figura14}. The inset in Fig. \ref{Figura12} shows that the probability of finding gaps larger than the particle size decreases with $ k $. As a consequence, when $k$ increases, patches are more likely to be trapped in a gap smaller than $ k $ and patches also need to diffuse over a smaller distance to bind to a particle adsorbed previously. Thus, as $ V_{m>k} $ goes to zero, the kinetics converges to RSA.

\begin{figure}[b!]
	\includegraphics[scale=0.5]{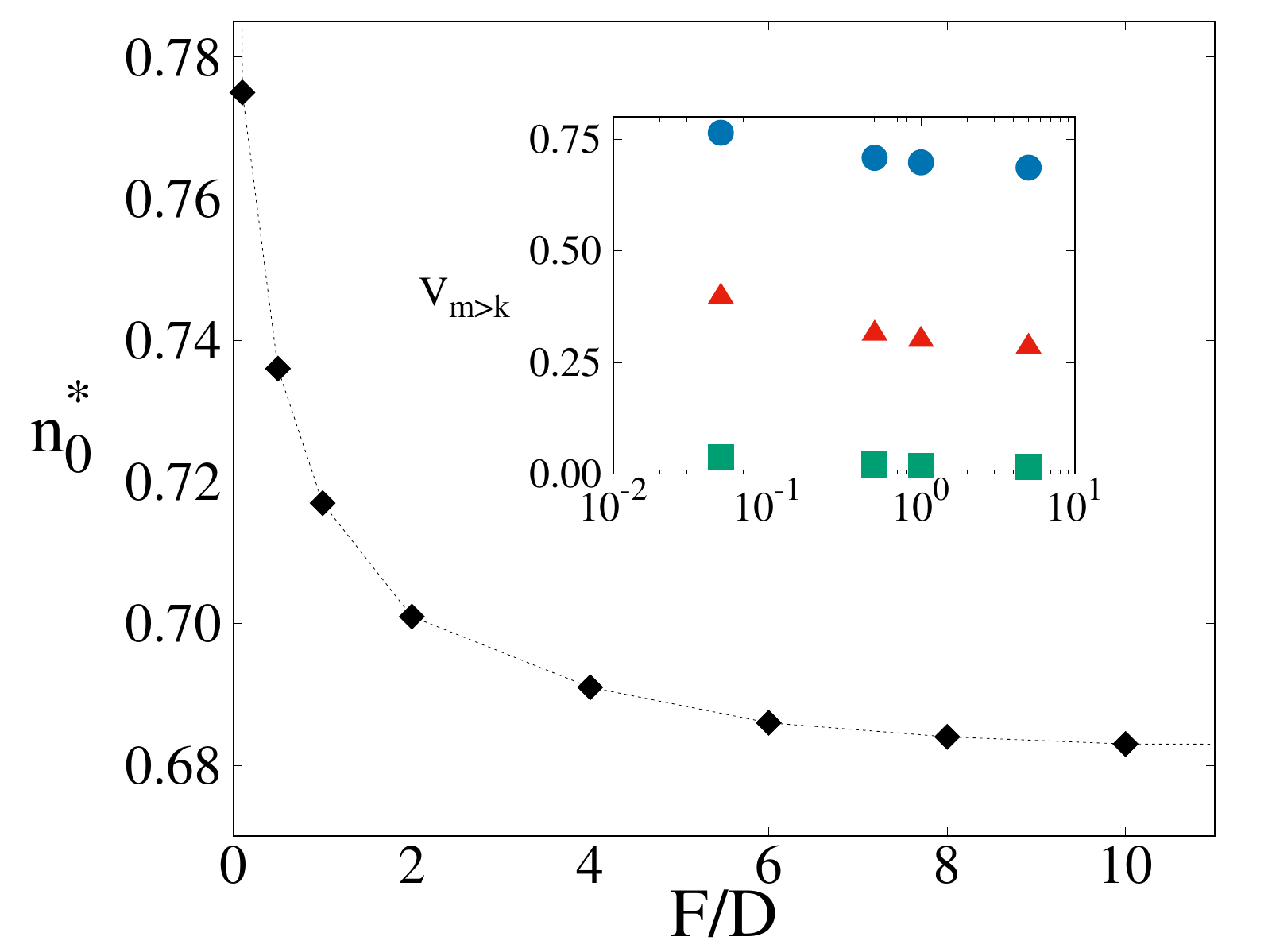}
	\caption{\label{Figura13} Plot of $ n_{0}^{*} $ as a function of $ F/D $. In the inset we have $ V_{m>k} $ as a function of $ F/D $, for $ n_{0}=\{0.1,0.3,0.7\} $ and $ k=2 $. In the inset the symbols and colors represent the same $ n_{0} $ as in Fig. \ref{Figura10}. Results are averages over $10^{4}$ samples of a one-dimensional lattice with $L=10^{6}$ sites.}
\end{figure}

To measure the crossover between our model and RSA as a function of $ F/D $, we introduce a new parameter $ n_{0}^{*} $. $ n_{0}^{*} $ is defined as the minimum value of $n_{0}$ for which the jammed-state coverage decreases with $k$. As shown in Fig. \ref{Figura13}, $ n_{0}^{*} $ decays with $ F/D $.

As $n_{0}$ increases, $ V_{m>k} $ converges to zero. For small values of $ F/D $, patches are more likely to find particles adsorbed previously before the next adsorption attempt. This favors the formation of gaps larger than $k$, as shown in the inset of Fig. 8. As $F/D$ increases, more adsorption events occur and the typical distance between particles is decreased. Thus $ n_{0}^{*} $ decreases.

\section{Conclusions}

We studied an extension of the Random Sequential Adsorption model (RSA), where particles can only adsorb on mobile patches. We found that, while the dynamics depends on the flux of particles $F$ and diffusion coefficient of the patches $D$, the jammed-state coverage solely depends on the ratio between the two. Supported by a mean-field calculation, we proposed that this feature results from the competition between two time scales: one related to adsorption and the other to diffusion. We revealed also a change in the functional dependence of the jammed-state coverage as a function of the particle size, depending on the density of patches.

Future studies might consider the effect of particle size dispersion or dimensionality of the substrate. For simplicity, we described patches as monomers. For larger patch sizes, more than one particle can adsorb on the same patch forming an aggregate. How does the size of the aggregates depend on the size of the particles and of the patches are questions of practical interest. 

\section{Acknowledgements}

We acknowledge financial support from the Portuguese Foundation for Science and Technology (FCT) under Contracts nos. UID/FIS/00618/2013 and SFRH/BD/131158/2017.

\bibliography{references}

\end{document}